# Regulating Trusted Autonomous Systems in Australia

Rachel Horne[1], Tom Putland[2] and Mark Brady [3]

[1] Trusted Autonomous Systems, Brisbane AUS, rachel.horne@tasdcrc.com.au, 0000-0002-1218-6872
[2] Trusted Autonomous Systems, Brisbane AUS, tom.putland@tasdcrc.com.au, 0000-0002-4260-1715
[3] Trusted Autonomous Systems, Brisbane AUS, mark.brady@tasdcrc.com.au, 0000-0002-5300-6225

**Abstract**
Australia is a leader in autonomous systems technology, particularly in the mining industry, borne from necessity in a geographically dispersed and complex natural environment. Increasingly advanced autonomous systems are becoming more prevalent in Australia, particularly as the safety, environmental and efficiency benefits become better understood, and the increasing sophistication of technology improves capability and availability. Increasing use of these systems, including in the maritime domain and air domain, is placing pressure on the national safety regulators, who must either continue to apply their traditional regulatory approach requiring exemptions to enable operation of emerging technology, or seize the opportunity to put in place an agile and adaptive approach better suited to the rapid developments of the 21$^{st}$ century.

In Australia the key national safety regulators have demonstrated an appetite for working with industry to facilitate innovation, but their limited resources mean progress is slow. There is a critical role to be played by third parties from industry, government, and academia who can work together to develop, test and publish new assurance and accreditation frameworks for trusted autonomous systems, and assist in the transition to an adaptive and agile regulatory philosophy. This is necessary to ensure the benefits of autonomous systems can be realised, without compromising safety.

This paper will identify the growing use cases for autonomous systems in Australia, in the maritime, air and land domains, assess the current regulatory framework, argue that Australia's regulatory approach needs to become more agile and anticipatory, and investigate how third-party projects could positively impact the assurance and accreditation process for autonomous systems in the future.

# 1 Introduction

Autonomous systems, broadly meaning systems capable of determining, initiating and executing action in pursuit of a goal, are being used around the globe to increase the safety and efficiency of tasks, and to lower economic and environmental cost.[1] *"To continue to make progress and innovate, ensuring that these systems have been designed responsibly and robustly will be key to safeguarding trust."*[2] While the technology enabling operation of these systems in the Australian maritime, air and land domains is rapidly advancing, the assurance and accreditation framework, which forms a necessary part of commercial operationalisation, is not keeping pace.

In Australia there is a gap in applied responsibility—it is within no specific Government department or organisation's remit to take a holistic approach to catering for emerging technology in an assurance and accreditation context. While the key national safety regulators have demonstrated an appetite for working with industry to facilitate innovation, limited resources limit progress. There is a critical role to be played by third parties from industry, government, and academia who can collaborate to develop, test and publish new assurance and accreditation frameworks for trusted autonomous systems, and provide recommendations for areas the regulators should focus on to ensure the benefits of autonomous systems can be realised, without compromising safety.

Recognising the lack of applied responsibility issue, and the benefits and commercial opportunities in facilitating the design, manufacture and use of autonomous systems, the Queensland State Government has funded a new groundbreaking activity. The Assurance of Autonomy Activity includes two projects – 'Enabling Agile Assurance of Drones in Queensland', and 'National Accreditation Support Facility Pathfinder (NASF-P) project'. The projects are intended to facilitate innovation within industry without compromising safety, by making it simpler and more efficient to design, manufacture, test, certify and operate trusted autonomous systems. Collaboration and an inter-disciplinary team, together with deep regulatory knowledge and strong stakeholder relationships will facilitate tangible results. This is a 'common good' activity, but will also bring commercial benefit, for example through increased business for Queensland test ranges and the creation of jobs. It is expected that this innovative project will have real impact over the next 24 months and beyond, and will result in a more efficient, faster assurance and accreditation process, without compromising safety.

This paper is presented in four parts: (1) Introduction, (2) Background, (3) Regulating Autonomous Systems, and (4) Conclusion. The Background includes a brief history, together with an overview of the key terms and use cases for each of the maritime, air and land domains. The Regulating Autonomous Systems section includes sub-sections on trust, regulatory philosophy, systems safety, and assurance and accreditation in Australia. The Conclusion reflects on the key messages presented in the paper, and points to the next steps. The paper is intended to consolidate cross-domain information on the uses of autonomous systems in Australia, provide thought leadership on the regulation of trusted autonomous systems, and provide an example of the Queensland Government's approach to addressing these complex issues.

# 2 Background

Australian industry has long been characterised by its ingenuity; borne of necessity due to the extremely remote and volatile environment that high risk operations such as land-based and deep-sea mining occurs in. Automated systems first came to Australia in automotive assembly plants and later to whitegoods manufacturers in the 1970s and 1980s, and included robots which were capable of highly accurate repetitive tasks within a set frame of movements.[3] Autonomous operation in mining came to Australia in the 1990s and has been in a state of continual development since then.[4] Australia now leads the world in the scale of autonomous mining operations.[5]

*"While automation has been a feature across industries in Australia for decades, the scope of tasks automated systems can complete has been enabled by the growth of digital, networked technologies and artificial intelligence. This transition – from primarily human operated to increasingly system or machine-operated systems – has necessitated new ways of thinking. Predicting and mitigating failures must take into account not only technical failures—such as errors in sensor measurements or system decision paths—but also human errors, both in robot design, and in interacting with robots."* [6]

Autonomous systems are becoming more prevalent in Australia, particularly as the safety, environmental and efficiency benefits become better understood, and the increasing sophistication of technology improves capability and availability. Increasing use of these systems, particularly in the maritime and air domain, is placing pressure on the national safety regulators, who must either continue to apply their traditional regulatory approach or seize the opportunity to put in place a more agile and adaptive approach better suited to the rapid developments of the 21st century.

*"As the breadth and application of semi-autonomous and autonomous systems continues to expand, the definitions of these terms evolve. There is no general consensus on these terms across disciplines. There are a few reasons for this: one being differing industry needs and approaches to describing and regulating systems nationally and internationally; the other being that some of these terms have histories and applications that predate robotics."*[7]

## 2.1 Maritime domain

There is no standardised taxonomy for how autonomous systems in the maritime domain are referred to. Some common generalised names include Autonomous Vessel (AV), Unmanned Vessel (UV), Maritime Autonomous Vehicle (MAV), Autonomous Vehicle (UxV), and Maritime Autonomous Surface Ship (MASS).[8] It is common for 'vessel' and 'vehicle' to be used interchangeably.[9] Other more specific names, indicating whether the vessel is a surface or subsurface vessel, are also used. These include Autonomous Surface Vessel (ASV), Autonomous Underwater Vessel (AUV), Unmanned Surface Vessel (USV), Unmanned Underwater Vessel (UUV), and Remotely Operated Vessel (ROV).[10]

In the maritime domain the primary uses for autonomous systems are currently scientific research, data gathering and hydrographic survey.[11] As technology develops and availability and capability increases, the use cases will continue to expand, for example into maritime surveillance and transport of goods. It is predicted that *"...Australia's maritime industry will increasingly feature more connected, intelligent and automated systems. The systems and infrastructure that supports high levels of automation and remote operation of vessels in Australia has the potential to significantly improve efficiency and safety for operators and seafarers alike."*[12]

Two of the primary drivers for use of autonomous systems in the maritime domain are the efficiency and cost-saving benefits that can be realised. For example, in the hydrographic survey industry, opportunities were identified early on to save a significant portion of fuel budgets by using small autonomous vessels to conduct the majority of surveying work. This economic incentive, which extends from fuel savings into the opportunity for concurrent operations without additional crew costs, and longer operation times, has driven the development of increasingly sophisticated autonomous vessels. In addition to economic incentives, autonomous systems have the capacity to remove humans from dangerous work circumstances and can reduce the environmental impact of commercial activities.

## 2.3 Air domain

Within Australia, the Civil Aviation Safety Authority (CASA) has adopted the terminology used by the International Civil Aviation Organisation (ICAO). This taxonomy broadly categorises Unmanned Aircraft Systems (UAS), i.e. those without pilots on-board the aircraft, as either autonomous aircraft or Remotely Piloted Aircraft Systems (RPAS). Previously UAS were referred to as Unmanned Aerial Vehicles (UAVs), but this terminology has fallen out of use. An autonomous aircraft is an unmanned aircraft *"…that does not allow pilot intervention during all stages of the flight of the aircraft"*.[13] This extreme definition is not entirely useful, although pervasive throughout aviation regulations. For the purposes of this paper, the term "autonomous" will not refer to the aviation definition, but instead align with the autonomous spectrum concept of "levels of autonomy" as used by Llyod's Register [14] (maritime), SAE [15] (automotive) and AIAA (aviation)[16]. The concept of autonomy applies to all types of aviation and is not restricted to unmanned aircraft specifically. However, for the purposes of this paper we will only consider unmanned aircraft (both autonomous aircraft and RPAS) incorporating some level of autonomy.

The current tasks undertaken by complex, autonomous UAS in Australia include survey/mapping tasks, in sectors such as mining, for example by Rio Tinto,[17] geographical and agricultural, for example Ninox Robotics,[18] and visual inspection-based tasks, for example by Telstra.[19] These activities will be joined by more complex, higher risk activities as the industry and regulatory context mature. Some of these tasks will include High Altitude Pseudo Satellite activities, for example by Airbus Defence & Space Zephyr, and Elson Space Engineering, package and cargo delivery, for example Wing [20] and Elroy Air [21], Urban Air Mobility (UAM), for example Joby Aviation S4 [22] and EmbraerX [23], and integrated emergency services operations, among many more activities.

## 2.4 Land domain

For the purposes of discussion around land-based autonomous systems, the key terminology used to refer to these systems are Unmanned Ground Vehicle (UGV), Unmanned Autonomous Ground Vehicle (UAGV), Autonomous Ground Vehicle (AGV), Autonomous Vehicle, Automated Vehicle or Unmanned Vehicle. These terms tend to be grouped together and referred to interchangeably depending on the circumstances.

In Australia, land-based autonomous systems are widespread, and use cases include diverse areas such as mining, which includes heavy mining equipment and smaller topological survey robots, agriculture, which includes heavy and light equipment and small-scale agricultural robots known as agribots. Emergency services incorporate search and rescue UGVs and legged robotics, as well as bomb disposal and firefighting drones. Heavy and light automated rail and some areas of road transport such as Connected Intelligent Transport Systems (CITS) may also be automated. Defence incorporates various types of automation such as radar and automated Close-In Weapons Systems (CIWS) and mobile automated platforms such as weapon carrying UGVs, or recovery vehicles and APCs. Other land-based autonomous systems include bio-mimetic systems, platform agnostic systems, and autonomous systems that are without physical instantiation such as Diagnostic AI, Legal Automation, µZero and other similar automated ML programs.

Autonomous systems are becoming more widely accepted as they have the potential to be more economical than traditional systems, and can remove humans from dirty, dangerous, or dull jobs, and in some areas – such as mining and manufacturing – are replacing humans altogether. The future landscape of autonomy will see industry, infrastructure, and society designed around autonomous systems rather than being added to or integrated subsequently and will likely be unrecognisable to people of today.[24] As the internet of things approaches autonomous systems in our everyday lives will become ubiquitous.

## 3    Regulating autonomous systems

*"Autonomous robots differ from most machines because of the computational components that lead to their intelligence and control. They also differ from most existing software systems because of their integration with physical machines. RAS-AI rely on observations perceived by sensors to decide how they should behave and control themselves, which is then translated to actuators, enabling robots to change behaviour, or state, and change in response to their physical environments."[25]*

There are many challenges associated with effectively regulating autonomous systems, including how to ensure trust in these systems, how to ensure the regulatory approach is the most appropriate option, how to adapt current systems safety approaches, and how to adapt current assurance and accreditation frameworks. The complex, interconnected nature of autonomous systems, including cybernetic systems, means that assurance as a concept must shift to account for the high levels of interdependency between core systems. For example, in an autonomous system some of the components that will require assurance include algorithms, software, hardware components, and the integrated robotic systems.[26]

*"In order for the robotics industry to thrive, a regulatory response that adapts, protects, engenders community trust, while accounting for rapidly evolving industrial environments, is essential. This is an area where investment can lead to improved outcomes for the whole robotics sector, and Australia can lead the world."[27]*

Trust is a key component of effective regulation, including from the perspectives of the regulator, regulated, industry and broader community. How trust is conceptualised, and how relevant considerations are included in regulatory frameworks, will change as increasing levels of autonomy are introduced.

## 3.1 Trust
*"Trustworthiness is a property of an agent or organisation that engenders trust in another agent or organisation."*[28] Building trust in autonomous systems is critical to uptake of the technology.[29]

Traditional technology is generally deemed safe if it complies with the regulatory framework set down and enforced by the Government or applicable regulator. For example, the public would be entitled to assume a commercial vessel is safe if it has met the survey, operational, and competency-based requirements imposed by the Australian Maritime Safety Authority. Trust is also supported by history of successful use – i.e., if a technology has been in use for one hundred years, there is a good understanding of the risks of failure and the results. Familiarity is another component of trust- if a technology has widespread usage and is visible in the media, trust will be engendered. A further component is trust by association – if an entity or person that is trusted uses or promotes use of a new technology, that can also engender trust.

Autonomous technology must be deemed safe in order to be accepted by prospective users, Government, the applicable regulator, and the public. This would be a simpler proposition if the Government or applicable regulator were experts in autonomous technology and had crafted well-tailored assurance and accreditation requirements, but that is rarely the case. The speed of technology development in the autonomous systems industry means that, early on, the only experts are the developers of the technology itself, and then the industry which is using it. In these circumstances industry-led projects to develop codes of practice or other regulatory content are vital, as they provide a baseline level of requirement to ensure safety among early-adopters, provide the Government or applicable regulator with a starting point for learning about the technology and then developing an appropriate regulatory approach, and they provide a level of confidence for members of the public.

An example of an industry-led collaboration is the Maritime UK Autonomous Systems Regulatory Working Group (MASRWG), which published a Code of Practice for Maritime Autonomous Ships in November 2017, and have provided an updated version each 12 months.[30] This Code provides a baseline set of requirements covering key issues such as design and manufacture, operation, skills and training, ship systems, cyber security, pilotage, and data recording.[31] While the Code is not a formal part of the UK maritime regulatory framework, it is used by industry as the accepted standard for autonomous vessels, and reflects current best practice.

A challenge for the Government or applicable regulator is to properly balance risk with regulatory overlay. If there is insufficient regulatory overlay to meet the risk of the operation, it will jeopardise trust in the technology and its regulation, but if the regulatory overlay is too heavy, it can stifle innovation and jeopardise trust in the intentions of the regulator. For this reason, adopting an appropriate regulatory approach is critical to ensuring innovation is supported, the benefits of autonomous systems are realised, and trust is maintained between all stakeholders involved.

## 3.2 Regulatory philosophy
There are numerous types of regulatory philosophy and approach, and the one adopted by specific countries for specific industries is generally determined by the predominant culture, traditions and corresponding expectations of government. While traditional regulatory approaches could be characterised as reactive, and are often based on prescription and inspections, more modern approaches move towards being proactive, agile, adaptive and anticipatory.

The concept of anticipatory regulation features these characteristics: inclusive and collaborative; future facing; proactive; iterative; outcomes based; and experimental at the forefront.[32] A key feature of anticipatory regulation is a focus on "co-design", where the regulator and industry work together to co-design standards and regulation that are fit for purpose and achieve the required outcomes. This approach focusses on the system and how to influence it. It introduces the concept of regulatory stewardship, whereby regulation is seen as an asset that helps things happen effectively.

In determining whether the predominant regulatory philosophy in Australia is appropriate, or if there is a need to change, it is necessary to consider ours, and the community's perspective on a number of questions: what is the role of a regulator in modern society; what are society's expectations; how do we understand and assess risk, and to what degree should regulators proactively ensure that risk it mitigated? The disruption to the Australian accommodation industry caused by Air BnB, and to the taxi industry caused by Uber, has shown that many people will place their trust in 5-star ratings systems, and in broad consumer confidence, even where there is far less Government or regulatory oversight over the house they have rented, or the car they are riding in, than was in place pre-disruption. Is it the case that regulators should draw back regulation a notch, and let market forces provide the level of safety consumers demand? In answering these core questions, we may find that a shift in regulatory philosophy is necessary.

We argue that Australia's regulatory approach must be reshaped to one that is agile, adaptive and anticipatory, and therefore capable of supporting fast updates of new technology, fast prototyping, and a paired back time to get new products to market. If Australian regulators could work towards becoming adaptive regulators, it would dramatically improve their ability to keep pace with technological change. The rise in popularity of human-centered design and the implementation of co-regulation are steps towards agile and anticipatory regulation, and will see Australia's ability to regulate emerging technology such as autonomous systems continue to increase.

### 3.3 Systems Safety approach
Regardless of regulatory approach, at some point there is a need to demonstrate that an operation is in compliance with a regulatory requirement. When approaching this task for complex systems, systems safety has become the de facto standard to demonstrate compliance. Systems safety is an engineering risk management process based on identifying and managing hazards in systems that balances safety and operability.[33]

Whilst systems safety focuses purely on systems, Safety Management Systems (SMSs) expand the concept beyond systems to organisations, people and processes. ICAO describes an SMS as, *"a systematic approach to managing safety, including the necessary organizational structures, accountabilities, policies and procedures"*.[34] Crucial to the SMS is organisational culture; the embodiment of safety thinking within each individual.[35]

Systems safety and SMSs are proven methods to drive better safety outcomes. However, as systems become increasingly autonomous, the accuracy of assumptions embedded in these practices (SMS and systems safety) are degraded. Additional assumptions for these autonomous systems, in the form of standards, guidance and practice do not fully exist and is an active area of research.[36]

There is a strong focus on System Safety and SMS as a way to ensure safety in both the maritime and air domains. For example, within the maritime regulatory framework, a SMS is required in order to satisfy General Safety Duties and operational requirements. In the aviation context, system safety is a design requirement for all type certified aircraft to ensure that the aircraft system is both reliable and resilient during operation. Safety Management Systems are required to be implemented by various approval holders in aviation in accordance with ICAO Standards and Recommended Practices (SARPs). This focus on Systems Safety and SMS will be an essential component in the successful assurance and accreditation of autonomous systems into the future.

### 3.4 Assurance and accreditation framework for autonomous systems in Australia
Autonomous systems in the maritime and air domains are regulated under an assurance and accreditation framework. Assurance is the process by which we validate that a vessel, aircraft, or discrete system complies with the applicable regulatory requirements and is therefore trustworthy. For example, part of the assurance process for an autonomous vessel would include a series of simulation-based testing, followed by physical testing (i.e. sea trials), together with the same series of inspections from an accredited marine surveyor that a traditional vessel is subject to. Accreditation refers to the certification required to be held by an autonomous system and its operator in order to comply with applicable regulatory requirements. For example, in the maritime domain, the standard certification requirements include a certificate of operation and a certificate of survey, in addition to specific certificate of competencies for the crew.

In Australia there is a lack of established assurance frameworks to integrate autonomous systems into traditional systems, which means neither industry, testing facilities, nor regulators, have clear or consistent expectations or understanding of what assurance activities are required to demonstrate compliance with requirements to indicate safe operations. This is inefficient and compromises trust, which makes it a priority to address.

The overarching System Safety and SMS approach does not need to be discarded, rather adaptions must be made in order to better accommodate autonomous systems. For example, the following components of the system safety process require adaption to accommodate autonomous systems:

1. An understanding of the acceptable level of safety for a system replacing traditionally human functions
2. The processes for decomposing safety-critical functional responsibilities and associated safety requirements between the human(s) and system
3. The methods to analyse the functional failure conditions for traditionally human functions, including methods to identify dependent failure mechanisms (for the purpose of designing for, or demonstrating independence)
4. The verification (4a) and validation (4b) methods to show compliance to derived safety requirements for novel techniques used to enable autonomous systems performing traditionally human functions. In particular this applies to machine learning elements of an autonomous system.

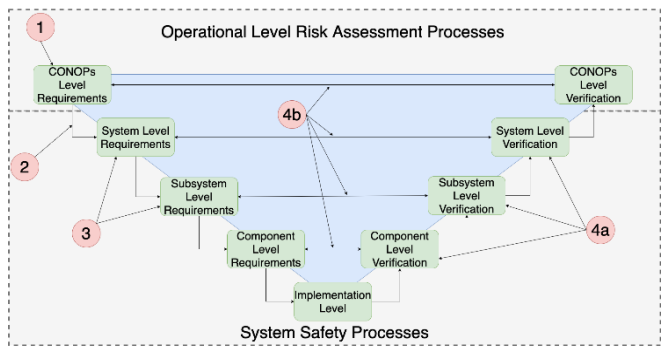

*Figure 1: Operational Level Risk Assessment Process (Tom Putland, TASD-CRC 2021)*

Accreditation typically relies on demonstrating compliance with regulatory requirements, including standards or defined codes of practice. However, when it comes to systems with autonomous capability, there are no tailored standards or codes of practice that relate to autonomy. In the air domain CASA uses a risk-based approach to approvals, which take into account the differences between autonomous aircraft and traditional aircraft. In the maritime domain, autonomous systems are subject to the same regulatory framework as traditional craft. This means autonomous operations generally rely on exemptions in order to operate, which take a long period of time to process and approve. There is often a lack of sufficiently sophisticated understanding of these autonomous systems, their capabilities and risks within the third-party surveying and inspection industry, and within regulators, which further slows down the accreditation process.

This lack of established standards or codes of practice for autonomous systems, and lack of sophisticated understanding within the surveying and inspection industry, and within regulators, represents a lost opportunity for the Australian economy to efficiently gain the economic benefit of new technology and to enable local industry to design, test and commercialise novel autonomous systems.

## 3.5 Regulatory requirements and issues in the maritime, air and land domains

Each of the maritime, air and land domains is subject to a different regulatory framework. In the maritime and air domains, regulation occurs at a Commonwealth level, under the remit of the Australian Maritime Safety Authority (AMSA) and the Civil Aviation Safety Authority (CASA) respectively. In the land domain regulation occurs primarily at a State level. This different jurisdictional approach means discussion of maritime and air domain issues and proposed solutions often occurs separately to discussions regarding the land domain.

*Maritime domain*
The Australian Maritime Safety Authority (AMSA) is a Commonwealth statutory authority responsible for maritime safety, protection of the marine environment from pollution, and search and rescue. AMSA regulates vessels operating within Australia's Exclusive Economic Zone (EEZ), including vessels capable of autonomous and remote-controlled operation.[37]

AMSA regulates three categories of vessels: domestic commercial vessels under the *Marine Safety (domestic commercial vessel) National Law Act 2012* (National Law Act), and regulated Australian vessels and foreign vessels under the *Navigation Act 2012* (Navigation Act). The legislation provides the overarching requirements, for example a vessel must be listed on a certificate of operation unless an exemption applies, Marine Orders provide the core regulatory content, for example the requirement to obtain a certificate of operation, and then standards such as the National Standard for Commercial Vessels (NSCV) provide detailed technical requirements. While neither the National Law Act nor the Navigation Act specifically refers to autonomous systems, the broad definition of "vessel" means autonomous systems are generally included.[38]

> *"The laws, Marine Orders, and standards that apply to all commercial vessels were written for traditional manned vessels, but remotely operated and autonomous vessels must also comply with them. As the unmanned vessels generally cannot comply with the design, construction, equipping and survey requirements applied to traditional vessels, and there are no tailored standards available to use, operators must seek exemptions in order to operate. This reliance on exemptions may not be feasible beyond the short term, due the administrative burden and delays it creates for operators and AMSA."*[39]

AMSA is actively working to improve their regulatory approach to autonomous systems, for example by seeking more hands-on experience [40], putting in place a policy to guide decision making [41], establishing an Autonomous Vessel team to triage queries and applications, working on a package of legislative amendments that would enable more flexibility to better address emerging technology [42], and establishing a partnership with the Trusted Autonomous Systems Defence Cooperative Research Centre (TAS) to explore issues around assurance and trust in autonomous systems.[43] AMSA has also affirmed their commitment to consulting and collaborating with industry, noting the opportunity it provides "…to gather information, test proposals for change, and ensure any proposed regulatory solutions are fit for purpose now and into the future."[44]

*Air domain*
Obligations under Article 8 to the Convention [45][46] requires a State to apply conditions on operations of unmanned aircraft systems (UAS) operated over their territory as long as the risk to civil aircraft is adequately addressed. Many nations, including Australia, have extended this concept to third parties on the ground under the concept of ensuring UAS operations are at least as safe as manned aviation. [47][48][49][50]

Regulation of UAS operations in Australia is therefore treated using a proportionate, risk-based framework [51] through Part 101 of the *Civil Aviation Safety Regulations 1998* (CASR). Operations outside of standard operating conditions (within visual line of sight, not over people, not near an aerodrome or above 400ft AGL) require multiple approvals for the various exceedances to the standard operating conditions.

Whilst CASA requirements are outcome based and therefore flexible and adaptable, there is currently a lack of direct guidance, processes or standards, particularly with respect to the design and demonstration of compliance with high level safety objectives for the highly automated or autonomous aspects of UAS. It is unlikely that regulators will have the capacity or expertise to detail specific standards or methods to design, test and build highly automated or autonomous systems. This will require a more collaborative approach with industry and academia, leveraging the capacity and expertise of industry.

A compounding effect that contributes to the aforementioned issues is the understaffing in both capacity and expertise that can be applied by CASA to effectively regulate this fast moving and novel aviation sector. This was identified in submissions to the Senate Standing Committee on Australia's General Aviation Industry in 2019.[52]

*Land domain*
The diverse applications of land-based autonomous systems fall under multiple regulatory areas. Each of the major areas where land-based autonomous systems operate have regulations specific to them, or some international safety standard that may apply. For example, land transport falls under the *Australian Vehicle Standards Rules 1999* (Cth)[53] and the *National Transport Commission (Road Transport Legislation— Vehicle Standards) Regulations 2006* (Cth)[54], and is based around the Australian Design Rules (ADR).[55] It is regulated in each jurisdiction with the assistance of peak advisory bodies such as Austroads and the National Transport Commission (NTC), which work to improve the regulatory process. Heavy vehicles are regulated separately under the *Heavy Vehicle (Vehicle Standards) National Regulation.[56]* Rail in Australia is administered by the Office of the National Rail Safety Regulator (ONRSR)[57] and each state and territory have legislation that mirrors ONRSR.[58] Agriculture, having no specific federal regulatory structure, is regulated in each state or territory under workplace health and safety regimes based around the *Work Health and Safety Act 2011* (Cth)[59] (WH&S), which have mirror legislation in each jurisdiction.

Each jurisdiction maintains its own guidance structure with codes of practice or industry guidelines that incorporate WH&S principles. Mining is regulated in each state or territory, and is covered by mineral resource legislation specific to the type of mining undertaken.[60] Automated mining is covered under various guidelines and safety standards,[61] applied by industry under an overarching safety framework based around WH&S principles that set workplace health and safety standards applicable to all mining. Emergency and disaster response services (EDS) are regulated separately in each state and territory under various regimes specific to the emergency response, such as fire,[62] police,[63] and ambulance,[64] – and as no specific autonomous technology regulations currently exist, these systems will likely fall under the broader remit of WH&S applicable to all branches of EDS.[65] Accordingly, where no controlling regulatory body at either the Commonwealth, state and territory level exists, the use of autonomous systems and technology will fall under the broad safety management requirements provided by the WH&S regimes in each jurisdiction. Where no assurance pathway for autonomous systems exists, there is a good case for developing a WH&S principle-based assurance pathway.

## 4     Opportunities for third party collaboration

There is a critical role to be played by third parties from industry, government, and academia who can work together to develop, test and publish a new assurance and accreditation framework for trusted autonomous systems, and provide recommendations for areas the regulators should focus on to ensure the benefits of autonomous systems can be realized, without compromising safety. This will address the cyclical problem that the development of improved technical standards is dependent on end user demand, but end user demand is dependent on the certainty that comes from having established standards and requirements in place.

In order to facilitate development and commercial operationalisation of autonomous systems, in some jurisdictions industry groups have collaborated to create guidelines and codes of practice, such as the Maritime Autonomous Surface Ships Industry Conduct Principles & Code of Practice [66] mentioned earlier. These documents, while not officially endorsed, provide industry and regulators with guidance on the minimum expectations for the build and safe operation of autonomous systems.

### 4.1 Initiative of the Queensland State Government: Assurance of Autonomy Activity

In Australia the key national safety regulators have demonstrated an appetite for working with industry to facilitate innovation, but their limited resources mean progress is slow. There is an opportunity for third parties, who sit outside of the traditional framework of applied responsibility, to take the lead in improving the assurance and accreditation of autonomous systems. This work is 'common good', in that it will have broad positive impact, but there is also commercial opportunity in it.

In Australia, this opportunity was identified by the Queensland State Government, and has resulted in funding for the Assurance of Autonomy Activity through the Trusted Autonomous Systems Defence Cooperative Research Centre (TAS). "*The Assurance of Autonomy activity aims to unlock Queensland's, and by extension Australia's, capacity for translating autonomous system innovation into operational capability, leveraging regulatory and technical expertise and strong stakeholder relationships to support industry and regulators. The Centre team, drawn from the Australian Maritime Safety Authority, Civil Aviation Safety Authority, and the University of Adelaide, have deep regulatory and technical expertise in autonomous systems, and bring a wealth of practical experience and strong stakeholder relationships to the project.*"[67]

The Assurance of Autonomy Activity currently hosts two projects. The first project, "Enabling Agile Assurance of Drones in Queensland", is led by Biarri Mathematical Consulting, supported by the Queensland University of Technology. It will deliver a smart tool to better connect operators and regulators, identify and demonstrate compliance with regulatory requirements, and possibly enable real-time approvals. This project responds to difficulties faced by industry in navigating the complexities of the regulatory framework, the resource burden faced by regulators, the fast-paced technological environment, the need for non-mariners such as scientists to engage in the regulatory process in a way not previously catered for, and the need to support development and operation of trusted autonomous systems. The second project, the "National Accreditation Support Facility Pathfinder (NASF-P) Project", is led directly by the Centre. It is intended *"...to improve the assurance and accreditation process for autonomous systems, support and promote Queensland test ranges, and pave the way for a new independent third-party entity that will offer world-class assurance support and consultancy services to domestic and international businesses. The activity will bring business to Queensland, and enhance its growing reputation as the Smart Drone State."[69]*

The projects have a collaborative focus, recognising the need for inter-disciplinary collaborations, demonstrated by strong relationships with domestic and international industry, academia, Government, and regulators, including the Australian Institute of Marine Science, Australian Maritime College Search at the University of Tasmania, and the Assuring Autonomy International Programme (AAIP) at the University of York.[68]

There are opportunities for third party collaborations in Australia, such as the Assurance of Autonomy Activity mentioned above, as well as the Australian Association for Unmanned System's new Maritime Working Group [70], to improve the assurance and accreditation approach for autonomous systems, and accelerate an improved regulatory approach which will facilitate innovation without compromising safety.

## 5  Conclusion

Australia is seeing a rapidly expanding market for autonomous systems in the maritime, air, and land domains, due to fast paced technological development, increasing availability and capability, and the safety, environmental and efficiency benefits they offer. We recommend a more agile, adaptive and anticipatory regulatory philosophy be implemented by regulators, with a focus on co-design, co-regulation, and trust, to ensure innovation is facilitated and the benefits of autonomous systems can be fully realized. Where Government and regulators are unable to lead innovation in an assurance and accreditation context, third party collaborations can address the gap and ensure progress is not stifled. The opportunities presented by rapidly developing autonomous systems technology will not be fully realized until an anticipatory regulatory philosophy is implemented, and an appropriately tailored assurance and accreditation framework is available for operators and regulators. Where these frameworks accurately reflect the community's expectations of Government and regulators, are appropriately tailored for application to autonomous systems, including balancing the regulatory overlay with the risk presented, trust will be engendered, and innovation accelerated.

*The research for this paper received funding support from the Queensland Government through Trusted Autonomous Systems (TAS), a Defence Cooperative Research Centre funded through the Commonwealth Next Generation Technologies Fund and the Queensland Government.*

**Reference list**
[1] Anastasio, G., Cole, J. and WF Smith, 'Framing Considerations of Autonomous Naval Ships for the Royal Australian Navy' (Conference Paper, International Maritime Conference, Pacific 2017).
[2] Devitt, K., Horne, R., Assaad, Z., Broad, E., Kurniawati, H., Cardier, B., Scott, A., Lazar, S., Gould, M., Adamson, C., Karl, C., Schrever, F., Keay, S., Tranter, K., Shellshear, E., Hunter, D., Brady, M., & Putland, T. (2021). Trust & Safety. *Robotics Roadmap for Australia V.2 [forthcoming]*. Robotics Australia Network.
[3], [4] Robotics has a long history in Australia, *Process and Control Engineering* (online 5 November 2013) https://pacetoday.com.au/robotics-has-a-long-history-in-australia/, last accessed 2021/02/11.
[5] Gleeson, D., 'Why the Pilbara Leads the World in Haul Truck Automation' International Mining (online 6 August 2019) https://im-mining.com/2019/08/06/pilbara-leads-way-haul-truck-automation/, last accessed 2021/02/10.
[6], [7] see 2.
[8] Horne R, Navigating to smoother regulatory waters for Australian commercial vessels capable of remote or autonomous operation: Stage 2 PhD document *[Forthcoming]* QUT, 2021/02/11.
[9] Rødseth, O. and Nordahl, H., Norweigan Forum for Autonomous Ships, *Definitions for Autonomous Merchant Ships*, (2017) http://nfas.autonomous-ship.org/resources/autonom-defs.pdf, ast accessed 2021/02/10.
[10] see 9.
[11] G Judson and R Horne, 'The regulatory approach for vessels capable of autonomous and remote-controlled operation' (Conference paper, International Maritime Conference, Pacific 2019).
[12] see 11.
[13] Regulation 101.097 of the Civil Aviation Safety Regulations 1998 (Cth).
[14] Lloyds Register, LR Code for Unmanned Marine Systems, 2017 ss. 4.1.2.
[15] Society of Automotive Engineers, SAE J3016 - Taxonomy and Definitions for Terms Related to Driving Automation Systems for On-Road Motor Vehicles.
[16] Anderson, E. Fannin, T. & Nelson, B. 2018, *Levels of Aviation Autonomy* (IEEE/AIAA 37th Digital Avionics Systems Conference 2018).
[17] Mining Global 2019, 'Rio Tinto enlists drone technology in Western Australia to drive efficiency and safety' Mining Global, https://www.riotinto.com/en/about/innovation/automation, last accessed 2021/02/15.
[18] Redrup, Y. 2016, 'Drones to become common in regional Australia: Ninox Robotics' <https://ninox-robotics.com/#introduction>, last accessed 2021/02/15.
[19] Wright, M. 2016, 'Eye in the sky watches over our mobile network', https://exchange.telstra.com.au/drone-pilot-gets-wings-tasmanian-telecommunications-first/, last accessed 2021/02/15.
[20] Martin L. 2019, 'Google's world-first drone delivery business wins approval in Canberra, accessed 15 February 2021', *The Guardian*, https://www.theguardian.com/australia-news/2019/apr/09/googles-world-first-drone-delivery-business-wins-approval-in-canberra, last accessed 2021/02/15.

[21] Chang, B. & Premack, R. 2019, 'Elroy Air has developed a transport drone it says can deliver anything from shipping cargo to humanitarian aid', Business Insider, https://www.businessinsider.com.au/elroy-air-autonomous-hybrid-aircraft-cargo-transport-2019-12?r=US&IR=T, last accessed 2021/02/15.

[22] Aviation Week, 'Joby Unveils EVTOL Design Details And Certification Plans' https://aviationweek.com/aerospace/urban-unmanned-aviation/joby-unveils-evtol-design-details-certification-plans, last accessed 2021/02/15.

[23] Reichmann, K. 2020, 'Australia Prepares for UAM with EmbraerX and Airservices CONOPS' https://www.aviationtoday.com/2020/12/21/australia-prepares-uam-embraerx-airservices-conops/, last accessed 2021/02/15.

[24] Brady, M., 'Is Australian Law Adaptable to Automated Vehicles' (2019) *Griffith Journal of Law and Human Dignity* 35-71.

[25], [26], [27] see 2.

[28] Devitt, K., Trustworthiness of Autonomous Systems, in H. A. Abbass et al. (eds.), *Foundations of Trusted Autonomy, Studies in Systems, Decision and Control* 117, https://doi.org/10.1007/978-3-319-64816-3.

[29] Australian Defence Magazine, ADF looks to build trust in autonomous systems, (online 5 November 2020) https://www.australiandefence.com.au/defence/unmanned/adf-looks-to-build-trust-in-autonomous-systems, last accessed 2021/02/15.

[30] Maritime UK, Maritime Autonomous Surface Ships Industry Conduct Principles & Code of Practice version 4, published 17 December 2020, https://www.maritimeuk.org/media-centre/publications/maritime-autonomous-surface-ships-industry-conduct-principles-code-practice-v4/, last accessed 2021/02/15.

[31] see 30.

[32] Anticipatory regulation, NESTA, https://www.nesta.org.uk/feature/innovation-methods/anticipatory-regulation/, last accessed 2021/02/11.

[33] According to NASA, "System Safety is the application of engineering and management principles, criteria and techniques to optimize safety within the constraints of operational effectiveness, time, and cost throughout all phases of the system life cycle": NASA, Office of Safety and Mission Assurance, 'System Safety' https://sma.nasa.gov/sma-disciplines/system-safety, last accessed 2021/02/15.

[34] Annex 19 to the Convention on International Civil Aviation, 2nd Edition, ICAO 2016.

[35] Sky Library, Safety Culture, https://www.skybrary.aero/index.php/Safety_Culture, last accessed 2021/02/15.

[36] see 2.

[37] see 11.

[38] Tethered remotely operated vessels (ROVs) are not considered 'vessels' as they are not considered to be 'capable of navigation'.

[39], [40] see 11.

[41] AMSA policy on regulation treatment of unmanned and/or autonomous vessels, published 20 February 2020, https://www.amsa.gov.au/vessels-operators/domestic-commercial-vessels/amsa-policy-regulatory-treatment-unmanned-andor#:~:text=AMSA%20will%20treat%20unmanned%20and,of%20relevant%20instruments%20as%20applicable, last accessed 2021/02/15.

[42], [43], [44] see 11.

[45] This interpretation of Article 8 "pilotless aircraft" to mean unmanned aircraft resides within the Global Air Traffic Management Operational Concept (Doc 9854), ICAO AN/458 1st Edition 2005.

[46] Article 8 to the Convention on International Civil Aviation (Doc 7300) ICAO ninth Edition 2006.

[47] For example, The European Union Aviation Safety Agency states this in (3) of the Cover Regulation to Implementing Regulation (EU) 2019/947 Commission Implementing Regulation (EU).

[48] In Australia this is covered through the *Civil Aviation Safety Regulations* 1998 (Cth) regulations 11.055 and 11.056.

[49] In the United Kingdom, this is covered under ss. 1.1.1 CAP 722.

[50] The Joint Authorities for Rulemaking on Unmanned Systems (JARUS), a conglomerate of over 60 national aviation authorities also contains this concept in AMC RPAS.1309 Safety Assessment of Remotely Piloted Aircraft Systems (JARUS 2015).

[51] See 'CASA takes risk-based approaches to regulatory action and decision making' (3) of CASA's Regulatory Philosophy https://www.casa.gov.au/about-us/who-we-are/our-regulatory-philosophy, last accessed 2021/02/12.

[52] Parliament of Australia, 'An inquiry into the current state of Australia's general aviation industry, with particular reference to aviation in rural, regional and remote Australia, https://www.aph.gov.au/Parliamentary_Business/Committees/Senate/Rural_and_Regional_Affairs_and_Transport/GeneralAviation/Submissions, last accessed 2021/02/11.

[53] *Australian Vehicle Standards Rules 1999* (Cth).

[54] *National Transport Commission (Road Transport Legislation—Vehicle Standards) Regulations 2006* (Cth).

[55] Link to ADR, https://www.infrastructure.gov.au/vehicles/design/index.aspx, last accessed 2021/02/12.

[56] *Heavy Vehicle (Vehicle Standards) National Regulation* (Cth).

[57] ONSR, https://www.onrsr.com.au/ last accessed 2021/02/10.

[58] For Rail Safety National Law as enacted in each jurisdiction, see, https://www.onrsr.com.au/about-onrsr/Rail-Safety-National-Law, last accessed 2021/02/15.

[59] *Work Health and Safety Act 2011* (Cth).

[60] For example, mining in Queensland is covered differently under the, *Mineral Resources Act 1989* (Qld); *Mining and Quarrying Safety and Health Act 1999* (Qld); *Mining and Quarrying Safety and Health Regulation 2001* (Qld); *Coal Mining Safety and Health Act 1999* (Qld); *Coal Mining Safety and Health Regulation 2001* (Qld).

[61] For example, relevant ISO Standards in relation to heavy autonomous mining equipment are: 17757 Earth-moving machinery and mining – Autonomous and semi-autonomous machine system safety; 19014 Earth-moving machinery – Functional safety – Part 1: Methodology to determine safety-related parts of the control system and performance requirements; and 16001 Earth-moving machinery – Object detection systems and visibility aids – Performance requirements and tests.
[62] In Queensland these fall under the *Fire and Emergency Services Act 1990* (Qld).
[63] In Queensland these fall under the *Police Service Administration Act 1990* (Qld).
[64] In Queensland these fall under the *Ambulance Service Act 1991* (Qld).
[65] In Queensland this is the *Work Health and Safety Act 2011* (Qld).
[66] See 30.
[67] Trusted Autonomous Systems Defence Cooperative Research Centre, Activities: Assurance of Autonomy, https://tasdcrc.com.au/activities/, last accessed 2021/02/11.
[68], [69] see 67.
[70] Australian Association for Unmanned Systems, Advisory Working Groups, https://aaus.org.au/advisory-working-groups/, last accessed 2021/02/11